\documentclass[aps,prb,twocolumn,floats,amsfonts]{revtex4}
\usepackage{dcolumn}
\usepackage{epsfig}
\begin{document}

\def\cm1{cm$^{-1}$}
\def\TiC{Ti$_8$C$_{12}~$}
\def\Td{T$_d~$}
\def\Th{T$_h~$}
\def\C3v{C$_{3v}~$}
\def\D2d{D$_{2d}~$}

\title {Predicted Infrared and Raman Spectra for Neutral 
 \TiC ~Isomers}

\author{ Tunna  Baruah $^{1,2}$,  Mark R. Pederson$^2$, 
M. L. Lyn$^3$ and A. W. Castleman Jr. $^3$}
\affiliation{$^1$Department of Physics,
    Georgetown University, Washington, DC 20057}
\affiliation{$^2$Center for Computational Materials Science,
   Code 6390, \\ Naval Research Laboratory, Washington, DC 20375}
   \thanks{Corresponding author: FAX: +1-202-404-7546;
           e-mail: pederson@dave.nrl.navy.mil} 
\affiliation{$^3$Departments of Chemistry and Physics,
    Pennsylvania State University, University Park, PA 16802   }

\date{\today}

\begin{abstract}
Using a density-functional based algorithm, the full IR and Raman 
spectra are calculated for the neutral \TiC cluster assuming 
geometries of \Th, \Td, \D2d and \C3v symmetry.  The \Th 
~pentagonal dodecahedron is found to be dynamically 
unstable.  The calculated properties of the relaxed structure having 
\C3v symmetry are found to be in excellent agreement with experimental 
gas phase infrared results,  ionization potential and electron affinity
 measurements.  Consequently, the results presented may be used as a 
reference for further experimental characterization using vibrational 
spectroscopy.
\end{abstract}

\maketitle

\section{Introduction}

In 1992, Castleman and co-workers
\cite{Guo1,Guo2} discovered the first of a family of magic number clusters, called 
metallocarbohedrenes, amidst a metal-carbon cluster distribution 
produced using a laser plasma reactor source.  Metallocarbohedrenes, 
or Met-Cars for short, are clusters of stoichiometry M$_8$C$_{12}$ 
(M = V, Zr, Hf, Ti, Nb, Mo, Fe, Cr) \cite{Guo1,Guo2,Wei,Pilgrim1} or 
Ti$_{8-x}$M$_x$C$_{12}$ (M = Zr, Hf, Nb, Mo, Ta, W, Si, Y, Nb, Mo, Ta, W)
\cite{Cartier1,Cartier2,Deng1}.  Bearing an unusual 2:3 metal-carbon 
reduced stoichiometric ratio, a pentagonal
 dodecahedral cage structure having \Th point group symmetry was 
proposed by Castleman to account  for ligand titration experiments
which suggested that the eight metal atoms had similar coordination
 \cite{Guo1}.  Shortly thereafter, Dance \cite{Dance1,Dance2} proposed 
a tetracapped tetrahedron of \Td symmetry and later showed that the \Th dodecahedron was higher 
in energy by  15 eV. Other geometries have also been proposed 
including structures belonging 
to the C$_{3v}$ point group \cite{Harris,Chen,Lin,Rohmer}.  
Comparison of the drift time of mass selected Ti$_8$C$_{12}^+$  
with calculated mobilities for various structures reinforces the 
proposed hollow cage structures instead of more closely packed 
cubic structures \cite{Bowers,Lee}.

Despite the considerable amount of interest 
in the chemical and physical properties of Met-Cars, the equilibrium 
geometry is still not known \cite{Rohmer}.
If these M$_8$C$_{12}$ clusters are to be realized as new materials, 
production and isolation in macroscopic quantities is imperative.  
Early experiments by Castleman {\it et al.} showed that Met-Cars
exposed to air were stable for short periods of time possibly due
to protection by soot \cite{Cartier3}.
  Most recently, Selvan and Pradeep also approached the bulk synthesis problem, using an 
arc-discharge \cite{Selvan}.  Infrared spectra of the soot 
after exposure to air over the course of 30 minutes showed the disappearance of a band at 
665 \cm1 and the emergence of others at 1457, 1122, and 872 \cm1.  The ``appearing'' 
bands were assigned to methylene wagging, C-C bond stretching, and methylene rocking modes, 
respectively.  As discussed below, the highest energy "appearing" mode (1457 \cm1) also 
is close to the energy range associated with Met-Cars.       
Time-lapse variations across IR spectra cannot be used to declare the presence of a "new" material 
when the disappearing or emerging peak(s) are not known or expected to be signatures of the 
species under investigation.

Recently, the infrared spectra for neutral \TiC, as well as for Ti$_8$C$_{11}$ and 
Ti$_{14}$C$_{13}$, 
have been measured in the gas phase by infrared multiphoton excitation using a pulsed linear 
accelerator \cite{Heijnsbergen}.  Though the measurement represents the vibrational signature for a free 
Met-Car, it serves as the first available experimental reference applicable for use in the 
identification of the presence of \TiC in unpurified samples.  
Recently,  Gueorguiev and  Pacheco \cite{Pacheco} have calculated the infrared
absorption spectrum for the \TiC. Their calculations  have shown the Met-Cars
with distorted \Td geometry to be the lowest energy ones while the experimental IR-absorption
spectrum compares better with the calculated spectrum for the \D2d structure. This
is clearly a finite temperature effect. However, the width of the experimental
peak near 1400 \cm1 is much larger than predicted by the above mentioned article. Neither  
the experimental\cite{Heijnsbergen} or theoretical\cite{Pacheco} works give strong evidence
that there is Met-Car infrared intensity near the 665 \cm1 range. Therefore an additional 
calculation  aimed at this question is useful. 
Another motivation of the present work is to identify the 
symmetry of the lowest energy structure within the density functional theory. A comparison  of the theoretical IR absorption spectra with
experiment will certainly be helpful in this regard. However, due
to the fact that the existing experimental measurements were carried out at
finite temperature, the characterization of the ground state
structure through IR measurement may be unreliable.
We show in the following that at finite temperature in which the experiments
are carried out,  the spectra  is a mixture of low lying isomers
leading to a large width of the high frequency peak.
In this paper, we present the complete vibrational modes for 
 a \D2d structure and a \C3v trigonal pyramid 
\TiC  as determined from all-electron density functional studies.   Apart from the 
IR absorption spectrum, we also present our calculated Raman scattering spectrum
for these structures. This is the first calculated Raman spectrum of \TiC. 
The reliability 
of our predicted Raman frequencies can be assessed based on agreement between 
the calculated IR spectrum and the gas phase IR results or other experimentally 
determined properties.

\section{	Computational Details}

The geometrical and electronic structures of \TiC molecule of different
structures were
studied using density functional theory \cite{Hohenberg}. The
calculations were performed with the NRLMOL code \cite{Nrlmol}
within the generalized gradient
approximation to the exchange-correlation energy \cite{PBE}.  The
Kohn-Sham orbitals were expanded in a basis of linear combination of
atomic orbitals where each orbital is expanded in a Gaussian basis set
centered on the atoms.
An optimized basis set \cite{Porezag_basis} was employed to speed
up the calculations.  The basis for the Ti atom consists of 
7 s-type, 5 p-type and 4 d-type orbitals each of which are 
constructed from a
linear combination of 19 Gaussians. For C atom, 5 s-type, 4 p-type and
3 d-type functions were used which in turn are composed of 12 
Gaussians. 
A variational integration mesh was used for the analytical computation 
of the potential \cite{mesh}.  Spin-polarized calculations  were performed 
taking all electrons into account.  The optimization of the clusters 
was symmetry adapted.  The massively parallel version of the
code \cite{parallel_nrlmol} was used and the calculations were carried out on a
cluster of PCs. 
The geometry optimizations were carried out by using the limited-memory
Broyden-Fletcher-Goldfarb-Shanno (LBFGS) scheme of minimization.
The geometry optimizations were carried till the forces  on the
atoms were below 0.001 a.u. For each geometry, the self-consistency
cycle was carried out till the energy differences were less than
0.000001 a.u. 
 
  To calculate the vibrational frequencies, a dynamical matrix
was constructed by first displacing the atoms by $\pm$ 0.08 a.u. 
  and calculating the  forces.
 The vibrational frequencies of the molecules were calculated by
a direct diagonalization of the dynamical matrix. The details of the
method can be found in Ref. \cite{Porezag}. The Raman and infrared
intensities  were obtained from the derivatives of the dipole
moment and the polarizability tensor 
which in turn
were obtained by applying an electric field
of strength 0.005 a.u.. For a high precision
calculation of the  vibrational frequencies and the Raman and IR
intensities,  the energies of the displaced geometries were converged
to a tolerance
of 1.0$\times$10$^{-8}$ a.u..

\section{	Results and discussion}

\begin{figure}
\epsfig{file=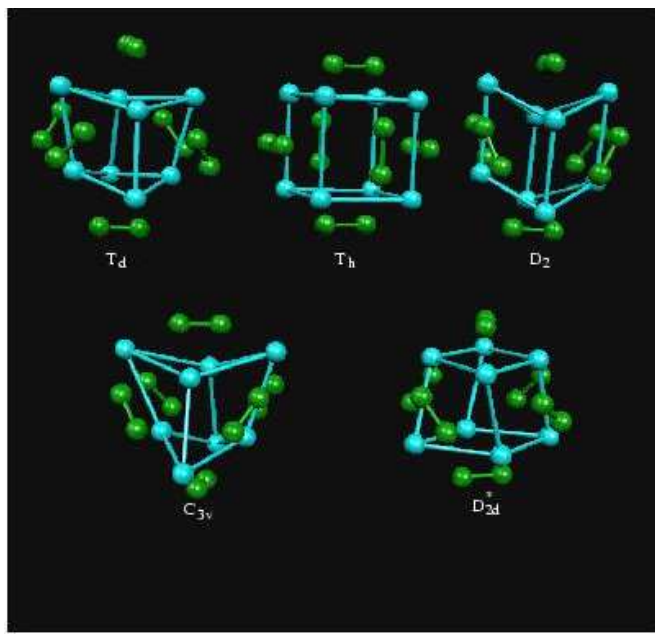,width=\linewidth,clip=true}
\caption{The optimized geometries of the \TiC with
various symmetries.
}
\label{geom}
\end{figure}

The earlier reports of the Met-Car  structure have shown the 
\TiC  clusters with a \Td geometry to be lower in energy
than the one with \Th symmetry.  
It may be recalled here that Dance \cite{Dance2} have shown a 
barrierless
reaction pathway between T$_h$ and the T$_d$ structures 
he considered.
The path was constrained with the D$_2$ symmetry. 
On the other hand,
Chen et al. \cite{Chen} have proposed a D$_{2d}$ structure to be 
more strongly bound. To obtain possible candidate structures for
the low energy isomers, 
 we have performed the geometry optimizations for  different
structures with \Td, \Th, \C3v, \D2d ,D$_{2h}$ and 
D$_2$ point group symmetries. 
Calculations for the lower symmetry structures 
albeit similar to the higher symmetry ones, were carried out so
as not to rule out any low symmetry structures. 
The lowest energy structures, which are unique 
are shown in Fig.  \ref{geom}. 
 We have carried out the optimizations for two structures with the 
\D2d symmetry in which the carbon dimers are oriented differently. 
We use a $*$ to distinguish between them. The one  labeled as \D2d has 
a structure
close to the \C3v structure and therefore is not shown in Fig. 
\ref{geom}. Similarly, we have not shown the structure of the D$_{2h}$
geometry since it is close to the \Th structure. Also, one of the
\Td ~structures, labeled as T$_d^*$ is close to the C$_{3v}$
and therefore is not shown.

The geometry optimization has revealed the
 structure with \C3v symmetry to be
the lowest energy structure among the structures considered here.  The 
energies of all the calculated
optimized structures relative to the energy of C$_{3v}$ structure
are given in Table \ref{table1}. We wish to
point out here that a similar structure with \Td symmetry (T$_d^*$)
was found to be slightly higher in energy (~0.13eV). 
The lower order of the  \C3v symmetry group removes some of the
degeneracies at the Fermi level.  A fully unsymmetrized
optimization leads  to a structure lower in energy by only 0.01 eV. 
Therefore, we accept the \C3v structure as the lowest energy
structure among the ones examined in this work.  We believe that
this structure is the same as the distorted \Td reported in 
Ref. \cite{Pacheco} .The T$_d$ and T$_h$ 
structures shown in Fig. \ref{geom} are significantly higher in energy,
 respectively, by 14.94 and 14.5 eV than the C$_{3v}$ structure. 
The energy difference between the T$_d$ and T$_h$ isomers is  
small (0.44 eV) .  The \D2d structure is also found to lie
higher in energy by 0.06 eV than the C$_{3v}$ structure. The other
structures with D$_{2d}^*$ and D$_2$ symmetry are still higher by $\sim$
2 eV.

To get more information about the stability of the
different conformers, we have calculated the 
atomization energies and the vertical ionization potential (vIP)
as well as the vertical electron affinities (vEA).
These values , along with the magnetic moments, 
and the HOMO-LUMO gaps
are summarized in Table \ref{table1}.  We found that while
the T$_h$   structure has magnetic moment  4 $\mu_{B}$, the other
structures are found to possess a smaller magnetic moment of
2 $\mu_B$. The
calculated binding or atomisation energy of all the different
structures are high which suggest that  these clusters are
highly stable. The vertical IP and EA are calculated by
assuming the structure of the charged cluster to be same as the
neutral one. The relative values of the vIP and vEA indicate the
relative stability of the clusters with different geometry.
Surprisingly, the higher energy isomers exhibit higher ionization 
potentials. 
The vertical IP of the \Td, \D2d and the \C3v structures 
range between 4.51-4.61 eV. 
The earlier reported values of calculated vIP for the T$_d$ structure
ranges between 4.37 - 4.7 eV  whereas the adiabatic IP is  4.43 eV 
\cite{Rohmer}. The vIP or IP  reported for other structures  
are considerably higher than the experimental value.
For the lowest energy structure with \C3v symmetry, we have
calculated the adiabatic IP. In this case, the
geometry of the charged cluster was allowed to relax. The
calculated ionization potential (4.47 eV)  is in excellent agreement
with the   recently 
determined value (4.4 $\pm$ 0.02 eV) from near 
threshold photoionization efficiency curves for 
the \TiC  \cite{Sakurai}.

The vertical electron affinity of the Ti  Met-Cars were  reported
for T$_d$ and T$_h$ structures. While the experimental vertical
electron affinity is 1.16 $\pm$ 0.05 eV, the adiabatic affinity of 
1.05 $\pm$ 0.05 eV \cite{Wang2} is lower. The vEA calculated in the present work
for the C$_{3v}$, D$_{2d}$ and T$_d^*$ range between 0.89 - 1.08 eV. The vEA for
other structures are relatively higher while the D$_{2h}$ shows a very
low vEA (Table I). 
The calculated adiabatic electron affinity
of the \C3v structure is 1.00 eV which is in excellent agreement with
the experimental value of 1.05 eV \cite{Wang2}. Apart from the C$_{3v}$ anionic 
structure,
we have optimized the geometries of the T$_d^*$ and the D$_{2d}$ 
structures also.
The anionic T$_d^*$ and C$_{3v}$ are degenerate with an energy 
difference of 
0.004 eV while the D$_{2d}$ anion lies 0.015 ev above.
The low electron affinity and the high ionization potential
of the \TiC signifies the low reactivity of this cluster.

 The energy ordering among the isomers and the good agreement with
experimental  IP and EA gives us
confidence that \C3v structure is a likely candidate for the lowest
energy structure within GGA.

  Another possible way of identifying the structure will be through
the infrared absorption spectrum.  Recently, Heijnsbergen {\it et al. } 
\cite{Heijnsbergen} have carried out the measurement of infrared 
resonance-enhanced multiphoton ionization spectrum   (IR-REMPI)
of the \TiC clusters within the frequency range of  400 cm$^{-1}$ to 
1600 cm$^{-1}$.  The IR-REMPI spectrum closely
resembles the conventional infrared absorption spectrum in peak
position and relative intensity.  This
experiment reports  a broad peak of the IR-REMPI spectrum centered
around 1395 cm$^{-1}$. 
A comparison of the calculated spectrum with the measured spectrum
can shed light on the possible candidate structures of the highly 
stable \TiC and also in identifying the Met-Cars in a mixed environment.
The vibrational spectrum of the high lying structures with T$_h$, 
T$_d$, D$_2$ and D$_{2h}$ symmetries have several imaginary 
frequencies which further indicates these structures to be highly 
unstable.  We, therefore, concentrate
on the analysis of the vibrational modes and the IR and Raman spectrum 
of the two lowest energy structures, namely, the C$_{3v}$ and D$_{2d}$ 
structures. 
Pacheco {\it et al.} \cite{Pacheco} have recently reported calculated
IR absorption spectrum
for the Met-Car with T$_d$, T$_h$ and \D2d structures and unsymmetric 
structures which are Jahn-Teller distorted. They have shown that
while the Jahn-Teller distorted T$_d$ structure is the lowest energy
structure, the absorption spectrum of the D$_{2d}$ structure matches
the experimental spectrum best. They suggest that at finite temperature,
the clusters with D$_{2d}$ symmetry are most abundant. 
The calculated spectra in Ref. \cite{Pacheco} were broadened by 
a Gaussian
of FWHM of 40 cm$^{-1}$ which merges the closely spaced peaks. 
For the sake of completeness, we also calculated the 
IR spectra for the \D2d structure reported in Ref. \cite{Pacheco}.
The starting geometry differs from the other D$_{2d}$ structure 
in the orientations of the carbon dimers. This structure,
upon optimization, distorts largely from the starting geometry. The
final structure is shown in Fig. \ref{geom} which we refer to as 
D$_{2d}^*$.  This structure, although high in energy with respect 
to the C$_{3v}$, is stable with no imaginary frequency associated 
with it.  However, the IR spectrum shows peaks at frequencies lower
than in the experimental spectrum - one at 1357 cm$^{-1}$ and a much
larger one at 1310 
cm$^{-1}$. Since this structure lies high in the energy scale,  we 
concentrate on the IR spectra of  the 
geometries labeled as C$_{3v}$ and the D$_{2d}$ .

\begin{figure}
\epsfig{file=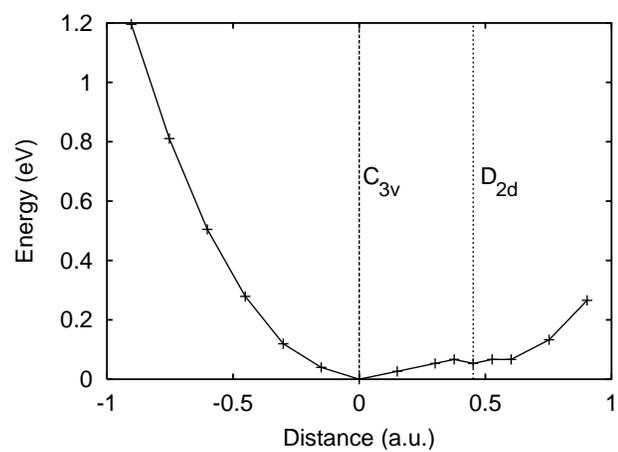,width=\linewidth,clip=true}
\caption{ A possible reaction path from the \D2d to  the \C3v
isomer. The x-axis shows the distances along the reaction path 
from the \C3v isomer.}
\label{c3v-d2d}
\end{figure}

Although both the C$_{3v}$ and the  D$_{2d}$ symmetry groups are 
subgroups of the T$_d$ symmetry group, they are independent. The  
small energy difference of 0.06 eV between them suggests the 
possibility of a symmetry breaking reaction path connecting the 
C$_{3v}$ and \D2d structures. Indeed, there exists such a path 
shown in Fig. \ref{c3v-d2d}.  The plot shows the small energy 
difference between the two structures.  However, an estimate of the 
vibrational frequencies for such an anharmonic potential
ruled out the possibility that at low or room temperature the
Met-Cars can vibrate between the two structures.  This fact again
establishes that at zero temperature, it is likely to conform to
 the \C3v structure.

\begin{figure}
\epsfig{file=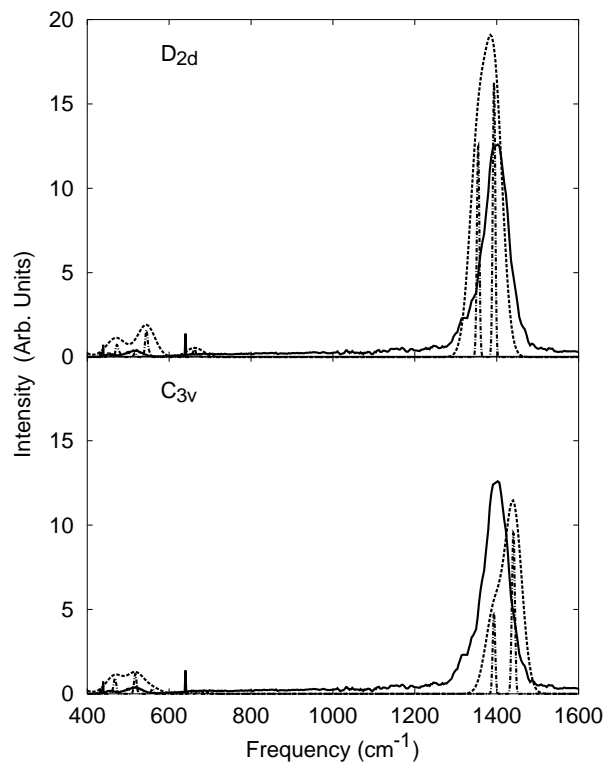,width=\linewidth,clip=true}
\caption{The calculated IR intensities for Ti$_8$C$_{12}$. The upper panel
shows the absorption spectra for D$_{2d}$ geometry and the lower one shows the
same for C$_{3v}$ geometry. The solid lines represent the experimental
spectra. The dash-dotted line shows the calculated spectra broadened with a 
gaussian of FWHM of 6 \cm1 while the dashed line shows the one broadened
with 40 \cm1. The convoluted curves are not renormalized. The 
theoretical intensities are in (D/$\AA$)$^2$amu$^{-1}$. The
experimental data are scaled down for comparison with theoretical data.
}
\label{IR}
\end{figure}

 The experimental spectrum was measured between 400 - 1600 cm$^{-1}$ 
range. Therefore, we present our calculated spectrum in this frequency 
range in Fig. \ref{IR}.  The spectra shown in Fig. \ref{IR} in the 
upper two panels were broadened with a Gaussian of full width half 
maximum (FWHM) of 6 \cm1 as well as 40 \cm1.  The  high resolution 
spectrum shows two clear peaks at 1393 \cm1 and 1442 \cm1 for the 
C$_{3v}$ structure (Fig. \ref{IR}, lower panel) while the \D2d structure
shows peaks at 1354 \cm1 and 1393 \cm1.  At temperature T=0 K,
both the structures have  strong absorption peaks near 1400 cm$^{-1}$.
The smearing of the peaks with a gaussian of FWHM of 40 cm$^{-1}$  
merges both the peaks and brings the spectra closer to the experimental
 one.  In the experimental spectra, we detect a shoulder near 1364 
cm$^{-1}$ which is clearly reproduced for the  
C$_{3v}$ spectra but at a higher frequency. Although the shape of the
IR absorption spectrum is correctly reproduced for the calculated 
\C3v spectra, the spectra is shifted slightly towards  high energy 
region. The peaks of the  \D2d spectra at 1354  and 1393 cm$^{-1}$ 
have comparable intensities and a smearing with Gaussian of FWMH of 
40 cm$^{-1}$ does not show a shoulder clearly and the broad peak is to 
the left of the experimentally observed peak.  In the low frequency 
region around 500 \cm1, the experimental spectrum shows  peaks at  455 
and  520 \cm1, which are correctly reproduced in case of \C3v structure
 and slightly shifted to higher frequency in case
\D2d structure. Based on this observation and as well as the fact that
the \C3v energy is lower than the \D2d, we believe that at temperature 
T=0K the IR spectrum will be dominated by that of the
C$_{3v}$ structure.  We have estimated the shift in the frequency of 
the highest peak of the \C3v structure due to the anharmonicity of 
the potential to be about 10 \cm1.  These estimates also showed that 
the lowest eight excitations would encompass a broadening of nearly 
30 \cm1.

\begin{figure}
\epsfig{file=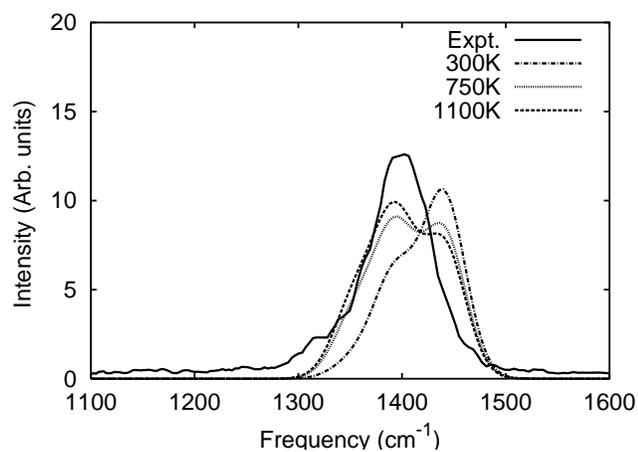,width=\linewidth,clip=true}
\caption{ The calculated IR intensities of the \TiC at 
different temperatures. The theoretical curves were convoluted
with Gaussian of FWHM of 40 \cm1.
}
\label{temp-IR}
\end{figure}

  Given that the experimental IR spectra was associated with clusters 
hot enough to undergo thermionic emission, the effect of thermal 
fluctuations will be strong in the experimentally
measured IR spectra. Therefore a statistically weighted spectra
of all the low-lying structure will better reproduce the experimental
IR absorption spectrum. In Fig. \ref{temp-IR}, we show the weighted
spectra of the C$_{3v}$ and the  \D2d structures which
are within a range of 0.06 eV. In the Fig. \ref{temp-IR}, we show the
calculated weighted  IR absorption for temperatures T=300K, 750K and 
1100K.  At room temperature, the spectra is dominated by the \C3v 
structure and weighted spectra has a peak shifted towards the high 
energy region. However, as
the temperature is increased, the contribution from the \D2d structure 
increases and the spectra shows two close peaks for 750 K.  At still 
higher temperature of 1100K, the peak due to the \C3v is reduced. The 
noticeable feature is that the width of the experimental peak is better
reproduced in the weighted spectra. The contribution from other isomers
will become important at still higher  temperature. Another noteworthy 
point is that the low energy region of the spectrum is better 
reproduced at room temperature.  

\begin{figure}
\epsfig{file=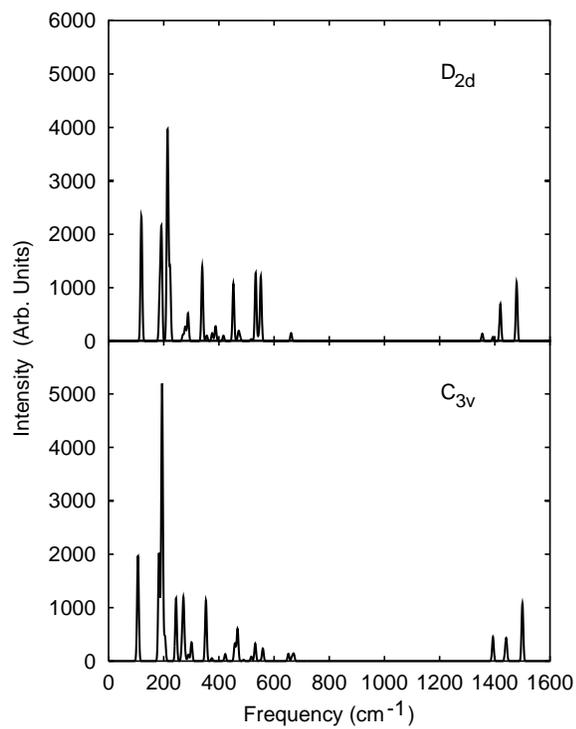,width=\linewidth,clip=true}
\caption{ The Raman scattering spectra of the low-lying D$_{2d}$ and 
the C$_{3v}$ structures. The spectra are broadened with a Gaussian
of 6 cm$^{-1}$.
}
\label{Raman}
\end{figure}

  For the purpose  of comparison with any probable future experimental 
investigation, we also present our calculated Raman scattering spectra 
for both \C3v and \D2d geometries. This is the first calculated Raman 
spectra for \TiC.  These plots are shown in Fig. \ref{Raman}. The Raman 
active modes are seen at lower frequencies which arises due to the 
Ti-Ti stretch mode and also twisting of the C-C  bonds.  These  modes 
will change the volume of the cluster and hence the polarizability 
which will  lead to  Raman activities.  The scattering 
intensities at high frequencies are low for both the structures. The 
spectra shown in Fig. \ref{Raman} has been convoluted with a Gaussian 
of FWHM 6 cm$^{-1}$.  A larger value of FWHM will smear out most of 
the fine structures seen in the plots. A few weak peaks are seen in 
the region around 1400 \cm1 where the IR activity is most strong. 
 In the Raman spectra of both the \C3v and \D2d structures, the most
prominent peaks occur in the region between 100 to 200 \cm1 which are
nearly similar (Fig. \ref{Raman}). Therefore,
Raman spectra in this low frequency region will not help in 
distinguishing the ground state structure between \C3v and \D2d. 
However, the spectra between 300 - 600 \cm1 are distinguishable as can 
be seen from Fig. \ref{Raman}. 
 A 
measurement of the Raman spectrum can greatly influence the debate over 
the equilibrium geometry of the Met-Cars.

\section{ Conclusions}

In conclusion, we have carried out extensive density functional 
calculations on the electronic structure and the vibrational states 
of the \TiC for different geometries. We find the geometry with \C3v 
symmetry to be the lowest energy structure. A study of the vibrational 
states show that the calculated IR absorption spectra for this geometry
 compares best with the experimental one. We find another structure 
with \D2d symmetry as a competing structure with an energy difference 
of only 0.06 eV. We find that a symmetry breaking reaction path exists 
from the \D2d to the \C3v structure. However, our estimates of the 
vibrational modes in the potential surface between the \C3v
and \D2d ruled out any possibility of cold clusters vibrating between 
the two closely placed geometries. We estimate the reaction barrier 
to be about 600K.  We point out that the experiments may be associated 
with hot clusters and show that the overall width of the high frequency 
IR spectra supports a high temperature mixture of \C3v and \D2d. The 
two peak character observed in our calculations could reduce to a  
shoulder structure with small perturbation of IR intensity and peak 
positions.  We also present our calculated  Raman spectra for these 
two low-lying structures. An experimental measurement of the Raman 
spectra may help in deciding the ground state geometry of the \TiC.

\section{Acknowledgements}
TB and MRP were supported in part by ONR grant N0001400WX2011.


\begin{table}
\caption{ The energies of the clusters relative to the lowest energy
structure, the atomisation energies (E$_b$), spin and the HOMO-LUMO
gap ($\Delta$) of the Ti$_8$C$_{12}$ clusters of various symmetries.
All values are in eV.}
 
\begin{tabular}[t]{lclllll} \hline
 
& Relative & E$_b$ & S & $\Delta$ & vIP & vEA    \\
 
& energy & \\
 
C$_{3v}$   &  0.00  & 7.08 & 1      & 0.12 & 4.61 & 0.89 \\
T$_h$      & 14.50  & 6.36 & 2      & 0.05 & 4.82 & 1.40 \\
T$_d$      & 14.94  & 6.34 & 1      & 0.22 & 5.68 & 2.09 \\
T$_d^*$    &  0.13  & 7.08 & 1      & 0.33 & 4.54 & 1.08 \\
D$_{2d}$   &  0.06  & 7.08 & 1      & 0.13 & 4.51 & 0.93 \\   
D$_{2d}^*$ &  1.78  & 7.00 & 1      & 0.44 & 5.35 & 1.35 \\
D$_{2h}$   & 14.47  & 6.36 & 1      & 0.25 & 6.26 & 0.66 \\   
D$_{2}$    &  1.99  & 6.99 & 1      & 0.07 & 5.18 & 1.53 \\   \hline

\end{tabular}
\label{table1}
\end{table}


\end{document}